\documentclass[pra,twocolumn,citeautoscript,superscriptaddress,longbibliography,pdftex]{revtex4-2}

\usepackage[hypertexnames=false]{hyperref}
\usepackage{grffile}
\usepackage{graphicx}
\usepackage{amsmath}
\usepackage{amssymb}
\usepackage[dvipsnames]{xcolor}
\usepackage{mathrsfs}
\usepackage{braket}
\usepackage[normalem]{ulem}
\renewcommand{\vec}{\mathbf}

\begin{document}

\title{DC transport in a dissipative superconducting quantum point contact}

\author{Anne-Maria Visuri}
\email{avisuri@uni-bonn.de}
\affiliation{Physikalisches Institut, University of Bonn, Nussallee 12, 53115 Bonn, Germany}
\author{Jeffrey Mohan}
\affiliation{Institute for Quantum Electronics, ETH Z\"urich, 8093 Z\"urich, Switzerland}
\author{Shun Uchino}
\affiliation{Advanced Science Research Center, Japan Atomic Energy Agency, Tokai 319-1195, Japan}
\author{Meng-Zi Huang}
\author{Tilman Esslinger}
\affiliation{Institute for Quantum Electronics, ETH Z\"urich, 8093 Z\"urich, Switzerland}
\author{Thierry Giamarchi}
\affiliation{Department of Quantum Matter Physics, University of Geneva, 24 quai Ernest-Ansermet, 1211 Geneva, Switzerland}

\begin{abstract}

We study the current-voltage characteristics of a superconducting junction with particle losses at the contacts. We adopt the Keldysh formalism to compute the steady-state current for varying transmission of the contact. In the low transmission regime, the dissipation leads to an enhancement of the current at low bias, a nonmonotonic dependence of current on dissipation, and the emergence of new structures in the current-voltage curves. The effect of dissipation by particle loss is found to be qualitatively different from that of a finite temperature and a finite inelastic scattering rate in the reservoirs.

\end{abstract}

\maketitle

\section{Introduction}

In open quantum systems, interesting physics arises from
the competition between the coherent Hamiltonian dynamics and the
incoherent, dissipative dynamics emerging from a coupling to an
environment~\cite{BreuerPetruccione2002, daley_quantum2014}. 
Such competition often occurs between first-order processes such as
single-site tunneling and particle loss on a lattice~\cite{YanYe2013,LabouvieOtt2016,TomitaTakahashi2017,FromlDiehl2019,FromlDiehl2020, WolffKollath2020,VisuriKollath2022,BenaryOtt2022,VisuriKollath2023}. However, there
are few studies on the fate of high-order coherent processes in the
presence of dissipation~\cite{groiseau_proposal_2021}. 
One would expect
that, because dissipation can destroy the coherence needed for 
such processes, they will be suppressed more quickly as the
order of the process increases. Such
high-order processes occur naturally in strongly correlated fermions, where a prominent example is the charge current in superconducting junctions under a subgap voltage bias. 
In such junctions, a phase bias leads to the Josephson supercurrent of Cooper pairs even in the absence of a voltage~\cite{Josephson1965,Josephson_supercurrents1965,likharev1986}. At small voltages, on the other hand, a DC current is generated by the simultaneous tunneling of a quasiparticle and multiple Cooper pairs. Due to the superfluid energy gap $\Delta$, quasiparticles cannot tunnel alone from one reservoir to another when the voltage $V$ is smaller than $2\Delta/e$, with $e$ the elementary charge. The energy gap can instead be overcome by multiple Andreev reflections (MAR)~\cite{KlapwijkTinkham1982, blonder_transition1982, averin_josephson1995, cuevas_superconducting1996, Bolech_point_contact2004, Bolech_Keldysh_study2005, SetiawanHofmann2022}, where effectively multiple Cooper pairs ($n_\mathrm{pair}\geq\Delta/\Delta\mu$) cotunnel, each providing the energy $2\Delta\mu$ \cite{cron_multiple-charge-quanta2001, cuevas_full2003}. The effect of dissipation on such a high-order process is an interesting question and yet little explored.

In addition to the supercurrent in superconducting junctions, there is therefore a normal current component generated by the quasiparticles transported in MAR~\cite{likharev1986}. Quasiparticle excitations are a central topic in the design of superconducting devices since they can lead to both decoherence and the dissipation of energy~\cite{LangMartinis2003, CatelaniPekola2022}. 
In solid-state superconductors, quasiparticle excitations typically dissipate their energy through inelastic scattering with phonons originating from lattice vibrations. 
The situation is slightly different in cold-atom experiments, where analogous systems of coupled fermionic superfluids have been realized~\cite{ValtolinaRoati2015, Husmann_quantum_point_contact2015, LuickMoritz2020, HuangEsslinger2023}. 
Since the phonon bath is absent, there is no similar energy dissipation mechanism for quasiparticles. In a cold-atom Josephson junction, dissipation was instead found to occur due to vortex creation~\cite{BurchiantiRoati2018,WlazlowskiMagierski2023} and the coupling between the condensate and Anderson-Bogoliubov modes~\cite{PhysRevLett.126.055301}. Specific dissipation mechanisms can furthermore be engineered in such experiments, so that the effects of dissipation on transport processes can be studied in a controlled way~\cite{corman_quantized_2019, lebrat_quantized2019}. The effects of particle losses as a dissipation mechanism have been studied in the case of bosonic superfluid transport~\cite{PhysRevLett.115.050601,LabouvieOtt2016,BenaryOtt2022} but less in the fermionic case. In a recent study~\cite{HuangEsslinger2023}, 
a spin-dependent dissipative beam was applied locally to a ballistic quantum point contact between two fermionic superfluid reservoirs, which made it possible to study the effect of local particle losses on the multiple Andreev reflection process. 

Theoretically, the effects of particle losses in fermionic superfluids have been studied in the context of orbital Feshbach resonances giving rise to two-body losses, modeled by either a non-Hermitian Hamiltonian~\cite{YamamotoKawakami2019} or within the Lindblad formalism~\cite{damanet_controlling_2019, damanet_reservoir_2019,YamamotoKawakami2021,MazzaSchiro2023}. 
In this work, we study the effects of particle losses in the transport setting, where we describe the dynamics by the Lindblad master equation. We focus on the DC current generated by a nonzero voltage across a dissipative quantum point contact (QPC) between two superconducting fermionic reservoirs. The dissipation occurs through particle losses acting locally at the contacts. The reservoirs are characterized by the superconducting gap, temperature, and a chemical potential which is in general different in the two reservoirs. We calculate the steady-state current through the QPC and analyze the effect of particle loss on the current-voltage characteristics. The particle loss is found to lead to an overall reduction of current but also an enhancement in the low-voltage regime where the current is generated by high-order multiple Andreev reflections. The local density of states of the superconductors is broadened by the dissipation, and we discuss how this broadening could lead to an enhancement of current. We also show that the effects of dissipation are distinct from those of inelastic scattering from phonons or a finite temperature.

Our model with particle losses described in the Lindblad formalism is natural in the context of cold-atom experiments where atoms are lost to the vacuum.
The results could also be relevant for solid-state superconducting junctions, where additional leads connected to the contact region can introduce local electron losses and gains~\cite{Morpurgo_tunable_supercurrent1998, Baselmans_reversing_supercurrent1999, MorpurgoKlapwijk2000, CrosserBirge2006}.
In this case, however, a multi-terminal description would be more appropriate since the temperature and chemical potential of all terminals plays a role.
The connection between the lossy system described by a master equation and a multi-terminal junction described in the Hamiltonian formalism is an interesting question. For noninteracting fermions, the analytic forms of the current were found to coincide when the loss term is replaced by a third terminal, and when certain conditions, such as the absence of gain from the third terminal, are satisfied~\cite{Uchino2022}.

In the following Sections, we discuss the model of the dissipative quantum point contact in Sec.~\ref{sec:model} and the theoretical method in Sec.~\ref{sec:keldysh}. Section~\ref{sec:keldysh} expands on the technical details already given in the supplemental material of Ref.~\cite{HuangEsslinger2023}. In Sec.~\ref{sec:high_transparency}, we analyze the current-voltage characteristics when the transparency of the contact is high, while Section~\ref{sec:weak_tunneling} focuses on the low-transparency regime. Further discussion and conclusions are presented in Sec.~\ref{sec:conclusions}.

\section{Model of the dissipative quantum point contact}
\label{sec:model}

\begin{figure}[h]
\includegraphics[width=0.55\linewidth]{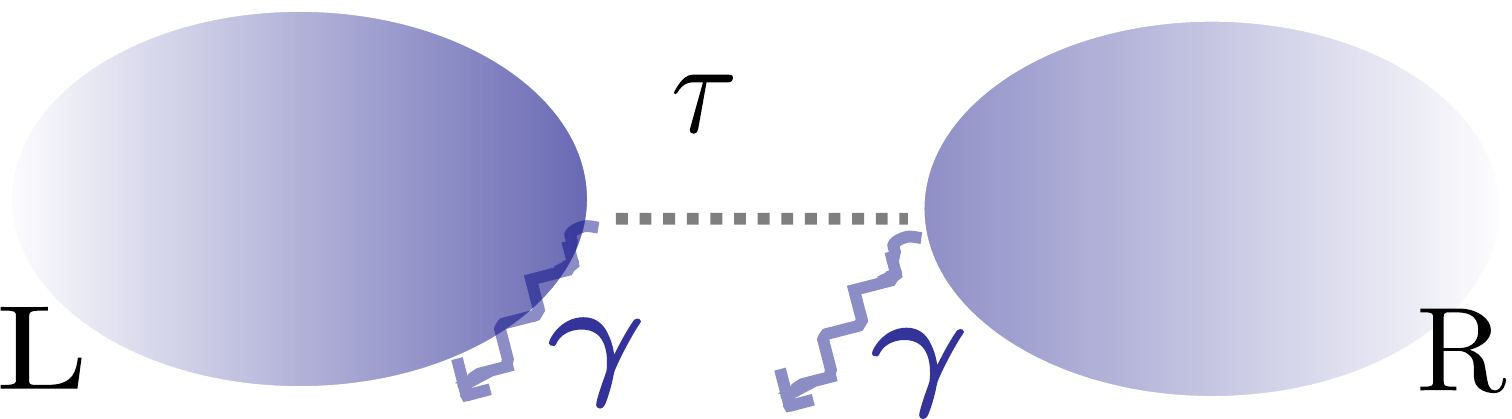}
\caption{\label{fig:schematic}
The two reservoirs are coupled by a quantum point contact, where particles tunnel directly with a tunneling amplitude $\tau$ between a single spatial point $\vec{r} = \vec{0}$ in each reservoir. In addition, the particle loss acts locally on these points. The loss rate is $\gamma$ for both reservoirs.}
\end{figure}

Our model of a dissipative junction is depicted in Fig.~\ref{fig:schematic}. Two reservoirs are coupled by tunneling between a single spatial point in each reservoir, representing the contact, and particle loss acts locally at these contacts with the same loss rate in both reservoirs. 
In the absence of dissipation, the system is described by the tunneling Hamiltonian
\begin{equation}
H = \sum_{i = L, R} H_i + H_{\text{tun}},
\label{eq:hamiltonian}
\end{equation}
where $H_{L, R}$ describes the left and right superconducting leads, respectively, and $H_{\text{tun}}$ describes the tunneling between them. The interaction in the superconducting leads is taken into account within a mean-field approximation. Thus,
the Hamiltonian in each superconducting lead is given by the following
quadratic form:
\begin{equation}
H_{i} = \sum_{\vec{k}} \Psi_{i \vec{k}}^{\dagger}\left[ \left( \epsilon_{\vec{k}} - \mu_i \right) \sigma_z + \Delta \sigma_x \right] \Psi_{i \vec{k}},
\label{eq:reservoir_hamiltonian}
\end{equation}
where $\epsilon_{\vec{k}}$ is the kinetic energy with momentum $\vec{k}$ and $\mu_i$ is the chemical potential in each lead. We adopt natural units with $\hbar = k_B = 1$. 
We consider a voltage $eV = \mu_L~-~\mu_R$ across the contact, and choose the chemical potentials symmetrically as $\mu_L = eV/2$ and $\mu_R = -eV/2$. The average chemical potential does not play a role here since the normal-state density of states is assumed constant. The superconducting energy gap is denoted by $\Delta$, which we consider real and independent of position, and $\sigma_{x, z}$ are Pauli spin matrices. We denote by $\Psi_{i \vec{k}}$, $\Psi_{i \vec{k}}^{\dagger}$ the Nambu spinors
\begin{equation}
\Psi_{i \vec{k}} = 
\begin{pmatrix}
\psi_{i, \vec{k} \uparrow} \\
\psi_{i, -\vec{k} \downarrow}^{\dagger}
\end{pmatrix}, \hspace{1cm}
\Psi_{i \vec{k}}^{\dagger} = 
\begin{pmatrix}
\psi_{i, \vec{k} \uparrow}^{\dagger} &\psi_{i, -\vec{k} \downarrow}
\end{pmatrix},
\end{equation}
where $\psi_{i, \vec{k} \sigma}^{\dagger}$ ($\psi_{i, \vec{k} \sigma}$) is the fermionic creation (annihilation) operator. The coupling between the reservoirs is described by $H_{\text{tun}}$, 
\begin{align}
H_{\text{tun}} &= -\tau \sum_{\sigma = \uparrow, \downarrow} [\psi^{\dagger}_{R \sigma}(\mathbf{0}) \psi_{L \sigma}(\mathbf{0}) + \text{H.c.}].
\label{eq:hamiltonian_tun}
\end{align}
Tunneling occurs between the points $\mathbf{r} = \mathbf{0}$ in each reservoir, and the tunneling amplitude is $\tau$.

In general, the density matrix $\hat{\rho}$
obeys a non-unitary evolution in the presence of dissipation. We treat such dynamics within
the Lindblad master equation
\begin{align}
\begin{split}
\frac{d\hat{\rho}}{dt} &= -i [H, \hat{\rho}] + \gamma \sum_{\sigma = \uparrow, \downarrow} \sum_{i = L, R}  \\
&\times \left[ \psi_{i \sigma}^{\phantom{\dagger}}(\mathbf{0}) \hat{\rho} \psi_{i \sigma}^{\dagger}(\mathbf{0}) - \frac{1}{2} \left\{ \psi_{i \sigma}^{\dagger}(\mathbf{0}) \psi_{i \sigma}^{\phantom{\dagger}}(\mathbf{0}), \hat{\rho} \right\} \right].
\end{split}
\label{eq:master_equation}
\end{align}
We consider a particle loss which acts locally on the ``contacts'' --  at one spatial point $\mathbf{r} = \mathbf{0}$ in each reservoir -- where tunneling also occurs.  The dissipation strength $\gamma$ is assumed to be independent of spin.

Transport is connected to the change in particle numbers of the reservoirs $\hat{N}_i = \hat{N}_{i \uparrow} + \hat{N}_{i \downarrow}$, where $\hat{N}_{i \sigma} = \int d\mathbf{r} \hat{\psi}_{i \sigma}^{\dagger}(\mathbf{r}) \hat{\psi}_{i \sigma}^{\phantom{\dagger}}(\mathbf{r})$. The time derivative of the particle numbers $\frac{d}{dt}  \braket{\hat{N}_i} = \frac{d}{dt} \text{Tr}\left[ \hat{N}_i \hat{\rho}(t) \right]$ is obtained from Eq.~(\ref{eq:master_equation}) and has two terms (see Appendix~\ref{app:currents})
\begin{align}
    \frac{d}{dt} \braket{\hat{N}_L} &= -I - I_L^{\text{loss}}, \\
    \frac{d}{dt} \braket{\hat{N}_R} &= I - I_R^{\text{loss}}.
\end{align}
The conserved current $I = \pm i \braket{[\hat{N}_i, \hat{H}]}$, where the different signs correspond to $i = L, R$, respectively, is the current which flows from the left to the right reservoir. It is expressed in terms of two-operator correlations as
\begin{equation}
    I = i \tau \sum_{\sigma = \uparrow, \downarrow} \left(\braket{\hat{\psi}_{R \sigma}^{\dagger}(\mathbf{0}) \hat{\psi}_{L \sigma}(\mathbf{0})} - \braket{\hat{\psi}_{L \sigma}^{\dagger}(\mathbf{0}) \hat{\psi}_{R \sigma}(\mathbf{0})} \right).
    \label{eq:conserved_current}
\end{equation}
The second term $I_i^{\text{loss}} = -\gamma \sum_{\sigma = \uparrow, \downarrow} \braket{\hat{n}_{i \sigma}(\mathbf{0})}$ describes the loss current from each reservoir and is proportional to the particle density $\braket{\hat{n}_{i \sigma}(\mathbf{0})} = \braket{\hat{\psi}_{i \sigma}^{\dagger}(\mathbf{0}) \hat{\psi}_{i \sigma}(\mathbf{0})}$ at the contacts. We focus here on the conserved current.

If both reservoirs are in the normal state, the conserved current is given by $I_N = \alpha G_0 V$, where $G_0$ denotes the conductance quantum $G_0 = 2e^2/h$. The transparency of the contact $\alpha$ depends on the tunneling amplitude and the dissipation,
\begin{equation}
    \alpha(\gamma) = \frac{4 \left(\frac{\tau}{W} \right)^2 \left( 1 + \frac{\gamma}{2 W} \right)}{\left[\left( 1 + \frac{\gamma}{2 W} \right)^2 + \left(\frac{\tau}{W} \right)^2 \right]^2},
\label{eq:transparency}
\end{equation}
and has values between 0 and 1. The value 1 corresponds to perfect transparency. The bandwidth $W = 1/(\pi \rho_0)$ is related to the normal-state density of states $\rho_0$ which we assume 
a constant determined at the Fermi energy. In the following, we report the tunneling amplitude $\tau$ and the dissipation rate $\gamma$ in units of $W$.

The model of a quantum point contact can be connected to that of a dissipative site coupled to two unitarily evolving reservoirs. The two models produce the same current-voltage characteristics in the limit of a high transparency $\alpha \approx 1$ of the quantum point contact and a large coupling between the reservoirs and the site, for both normal and superconducting reservoirs. The model of a dissipative site was employed to describe a lossy quantum point contact in Ref.~\cite{HuangEsslinger2023}.

\section{Keldysh formalism}
\label{sec:keldysh}

To calculate the nonequilibrium expectation values in Eq.~(\ref{eq:conserved_current}), we use the Keldysh formalism~\cite{Kamenev_field_theory2011, Sieberer_Keldysh2016}, where such expectation values are given by Keldysh Green's functions. In the path integral formulation, expectation values are calculated as path integrals over a closed time contour. The integration is performed by introducing the Grassman variables $\psi = (\psi^+, \psi^-)$ for the forward and backward time branches. For convenience, we apply the Keldysh rotation~\cite{Kamenev_field_theory2011}:
\begin{equation}
    \begin{pmatrix}
        \psi^1\\
        \psi^2
    \end{pmatrix}    
    =\frac{1}{\sqrt{2}}\begin{pmatrix}
        1 & 1\\
        1 & -1
    \end{pmatrix}
    \begin{pmatrix}
        \psi^+\\
        \psi^-
    \end{pmatrix},
\end{equation}
written in the bosonic convention, which allows to express correlation functions in terms of the advanced, retarded, and Keldysh Green's functions.
The partition function is written as
\begin{equation}
\mathcal{Z} = \int \mathcal{D}[\psi, \bar{\psi}] e^{i \mathcal{S}[\psi, \bar{\psi}]},
\end{equation}
where $\mathcal{S}$ is the Keldysh action. The master equation~(\ref{eq:master_equation}) can be mapped onto an action which is a sum of the coherent and dissipative parts~\cite{Sieberer_Keldysh2016}: $S = S_L + S_R + S_t + S_{\text{loss}}$.
Here, the coherent terms $S_L$, $S_R$, and $S_t$ arise from the Hamiltonian~(\ref{eq:hamiltonian}) and $S_{\text{loss}}$ from the dissipative term in Eq.~(\ref{eq:master_equation}). 
Two-operator correlation functions are calculated as Gaussian path integrals
\begin{equation}
\braket{\psi^a \bar{\psi}^b} = \int \mathcal{D}[\bar{\psi}, \psi] \psi^a \bar{\psi}^{b} e^{i S[\bar{\psi}, \psi]} = i \mathcal{G}_{a b},
\label{eq:correlation_function}
\end{equation}
where $a, b$ denote the sets of relevant indices $(1, 2, i, \sigma)$ and $\mathcal{G}_{a b}$ denotes the matrix element of the system Green's function as discussed in the following. As we are interested in steady-state observables, it is convenient to use the frequency basis. The relevant expectation values only depend on $\psi$ and $\bar{\psi}$ at position $\mathbf{r} = \mathbf{0}$, and in the following, we denote $\psi(\omega) = \psi(\mathbf{r} = \mathbf{0}, \omega) = \sum_k \psi(k, \omega)$. The conserved current of Eq.~(\ref{eq:conserved_current}) is given by
\begin{equation}
I = \frac{i \tau}{2} \sum_{\sigma = \uparrow, \downarrow} \int \frac{d \omega}{2 \pi} \left( \braket{\psi_{R \sigma}^1 \bar{\psi}_{L \sigma}^1} - \braket{\psi_{L \sigma}^1 \bar{\psi}_{R \sigma}^1} \right).
\end{equation}

\subsection{Keldysh action for uncoupled leads}

We take advantage of the matrix representation of the action
\begin{equation}
S = \int \frac{d \omega}{2 \pi} \bar{\mathbf{\Psi}}(\omega) \mathcal{G}^{-1}(\omega) \mathbf{\Psi}(\omega).
\label{eq:general_action}
\end{equation}
For uncoupled leads, $\mathbf{\Psi}$ is the four-component Nambu-Keldysh spinor 
\begin{equation}
\mathbf{\Psi}_i = 
\begin{pmatrix}
\psi_{i \uparrow}^1	&\bar{\psi}_{i \downarrow}^1	&\psi_{i \uparrow}^2	&\bar{\psi}_{i \downarrow}^2
\end{pmatrix}^T
\end{equation}
with $i = L, R$. We denote by $\bar{\mathbf{\Psi}}$ the corresponding row vector of the bar fields. The inverse Green's function $\mathcal{G}^{-1}$ has the structure
\begin{equation}
\mathcal{G}^{-1} = 
\begin{pmatrix}
0	&\left[ g^\mathcal{A} \right]^{-1} \\
\left[ g^\mathcal{R} \right]^{-1}	&\left[ \mathcal{G}^{-1} \right]^{\mathcal{K}}
\end{pmatrix},
\label{eq:general_inverse_greens_function}
\end{equation}
where $\left[ \mathcal{G}^{-1}\right]^{\mathcal{K}} = -\left[ g^{\mathcal{R}} \right]^{-1} g^{\mathcal{K}} \left[ g^{\mathcal{A}} \right]^{-1}$, and $g^\mathcal{A}$, $g^\mathcal{R}$, and $g^\mathcal{K}$ are the advanced, retarded, and Keldysh components of the Green's function. 
We use here the boson notation for fermionic coherent states rather that the one of Larkin and Ovchinnikov. In the case of no dissipation, the uncoupled reservoirs are at thermodynamic equilibrium, and the Keldysh component $g^\mathcal{K}$ is determined by the fluctuation-dissipation relation as~\cite{Kamenev_field_theory2011}
\begin{equation}
g^\mathcal{K}(\omega) = \left[ g^\mathcal{R}(\omega) - g^\mathcal{A}(\omega) \right][1 - 2 n_F(\omega)].
\end{equation}
Here, $n_F$ denotes the Fermi-Dirac distribution $n_F(\omega) = \left( e^{\omega/T} + 1\right)^{-1}$ with temperature $T$. The retarded and advanced Green's functions generally contain information about the excitation spectrum, while the Keldysh Green's function is connected to correlations and the occupation of modes.

For fermions with spin, $g^\mathcal{A}$, $g^\mathcal{R}$, and $g^\mathcal{K}$ are matrices of size $2 \times 2$, and the matrix elements can be computed from the Hamiltonian of Eq.~(\ref{eq:reservoir_hamiltonian}). The inverse of the retarded and advanced component is given by
\begin{equation}
[g^{\mathcal{R}, \mathcal{A}}(\omega)]^{-1} = \frac{W}{\sqrt{\Delta^2 - (\omega \pm i\eta)^2}} 
\begin{pmatrix}
\omega \pm i\eta	&\Delta	\\
\Delta				&\omega \pm i\eta
\end{pmatrix},
\label{eq:retarded_advanced}
\end{equation}
where $\eta > 0$ is an infinitesimal constant which regularizes the Green's functions.
Introducing local particle losses in the reservoirs, as written in Eq.~(\ref{eq:master_equation}), modifies the the reservoir Green's functions. 
As explained in Appendix~\ref{app:Keldysh_action}, we obtain $[g_i^{\mathcal{R}}]^{-1}$ and $[g_i^{\mathcal{A}}]^{-1}$ at $\mathbf{r} = \mathbf{0}$ as
\begin{align}
    [g_i^{\mathcal{R}, \mathcal{A}}]^{-1} = [g_i^{\mathcal{R}, \mathcal{A}}]_{\gamma = 0}^{-1}
    \pm \frac{i}{2}
    \begin{pmatrix}
        \gamma	&0	\\
        0	&\gamma
    \end{pmatrix}
\end{align}
and
\begin{align}
[G_i^{-1}]^{\mathcal{K}} = [G_i^{-1}]_{\gamma = 0}^{\mathcal{K}} +i
\begin{pmatrix}
     \gamma	&0	\\
    0	&- \gamma
\end{pmatrix},
\label{eq:keldysh_block_dissipation}
\end{align}
where $[G_i^{-1}]_{\gamma = 0}^{\mathcal{K}} = \left( [g_i^{\mathcal{R}}]_{\gamma = 0}^{-1} - [g_i^{\mathcal{A}}]_{\gamma = 0}^{-1} \right) [1 - 2 n_F(\omega)]$ and $[g^{\mathcal{R}}]_{\gamma = 0}^{-1}$ and $[g^{\mathcal{A}}]_{\gamma = 0}^{-1}$ are given by Eq.~(\ref{eq:retarded_advanced}).

While we generally set the constant $\eta$ infinitesimal in Eq.~(\ref{eq:retarded_advanced}), a finite value of $\eta$ can be used to model the damping of the quasiparticle excitations by inelastic scattering from phonons in the superconducting leads~\cite{cuevas_superconducting1996}. In a closed system ($\gamma = 0$) with a finite damping rate, the particle number is conserved, while in the open quantum system with a particle loss, there is no particle number conservation. This difference is reflected in the form of the Keldysh Green's function, as seen in Equation~(\ref{eq:keldysh_block_dissipation}): In the absence of particle loss, $[\mathcal{G}_i^{-1}]^{\mathcal{K}}_{\gamma = 0}$ is proportional to $1 - 2n_F(\omega)$, while in the presence of loss ($\gamma > 0$, $\eta \to 0$) it is independent of frequency. In Secs.~\ref{sec:high_transparency} and~\ref{sec:weak_tunneling}, we compare the effects of a finite $\eta$ on the current-voltage characteristics to those of a particle loss.

\subsection{Superconductor-superconductor junction}
\label{sec:superconductor-superconductor}

To describe a superconducting contact, one can use a
leading-order perturbative analysis in the limit of small $\tau$ where the contribution of higher-order tunneling processes to the current are negligible~\cite{blonder_transition1982}. However, for $\tau/W \to 1$, one needs to take into account all orders of tunneling. This can be done conveniently in the Keldysh formalism. The action describing the superconducting contact is formed by coupling the left and right matrix blocks, given by Eqs.~(\ref{eq:general_inverse_greens_function})--(\ref{eq:keldysh_block_dissipation}), by tunneling matrix elements. 
In the absence of a chemical potential bias, the action including the left and right reservoirs, tunneling, and loss could be represented by an $8 \times 8$ matrix $\mathcal{G}^{-1}$. One must however take into account that the interaction couples opposite-spin fermions within each lead at frequencies which have an equal but opposite-sign shift from the chemical potential $\mu_{L, R}$ of each lead. 
When there is a potential bias, this reference level is different in the two leads, while the tunneling matrix elements connect states at the same absolute frequency. This leads to an action which is not block diagonal in frequency. In fact, the interaction and tunneling terms connect matrix elements with increasing frequencies which follow the recursion relation~\cite{Bolech_point_contact2004, Bolech_Keldysh_study2005}
\begin{align}
\begin{split}
&\omega_n = 2 \mu_{2 - n \text{mod} 2} - \omega_{n - 1} \\
&\omega_{-n} = 2 \mu_{1 + n \text{mod} 2} - \omega_{1 - n}.
\end{split}
\label{eq:recursion}
\end{align}
Here, $\mu_{1, 2} = \mu_{L, R}$ and $n = 1, 2, \dots$ 
The initial frequency $\omega_0$ is the one of the spin-up fermion in the left reservoir, $\omega_1$ is the frequency of spin down in the left reservoir which is coupled to $\omega_0$ by the interaction term. As the spin-up fermion at frequency $\omega_0$ tunnels to the right reservoir, it is coupled by the interaction to a spin-down fermion in the right reservoir at frequency $\omega_{-1}$, and so forth. 
The action is therefore represented by a matrix of infinite size, as detailed in Appendix~\ref{app:infinite_matrix}. 

Physically, the coupling of matrix elements at increasing frequencies amounts to pair cotunneling, \textit{i.e.}, multiple Andreev reflections.
One can see from Eq.~(\ref{eq:retarded_advanced}) that the matrix elements which contain the superconducting gap decay with frequency. The matrix therefore approaches a block-diagonal form at $|\omega - \mu_{L, R}| \to \infty$, and it can be truncated at a certain size, corresponding to a certain order of the tunneling process. We compute expectation values by numerically inverting the matrix $\mathcal{G}^{-1}$~\cite{Bolech_point_contact2004, Bolech_Keldysh_study2005,Husmann_quantum_point_contact2015,yao_controlled_2018}, rather than solving the Keldysh Green's functions from the Dyson equation~\cite{Uchino2022}.

\section{High transparency of the contact}
\label{sec:high_transparency}

\subsection{Current-voltage characteristics}

\begin{figure}[h]
\includegraphics[width=\linewidth]{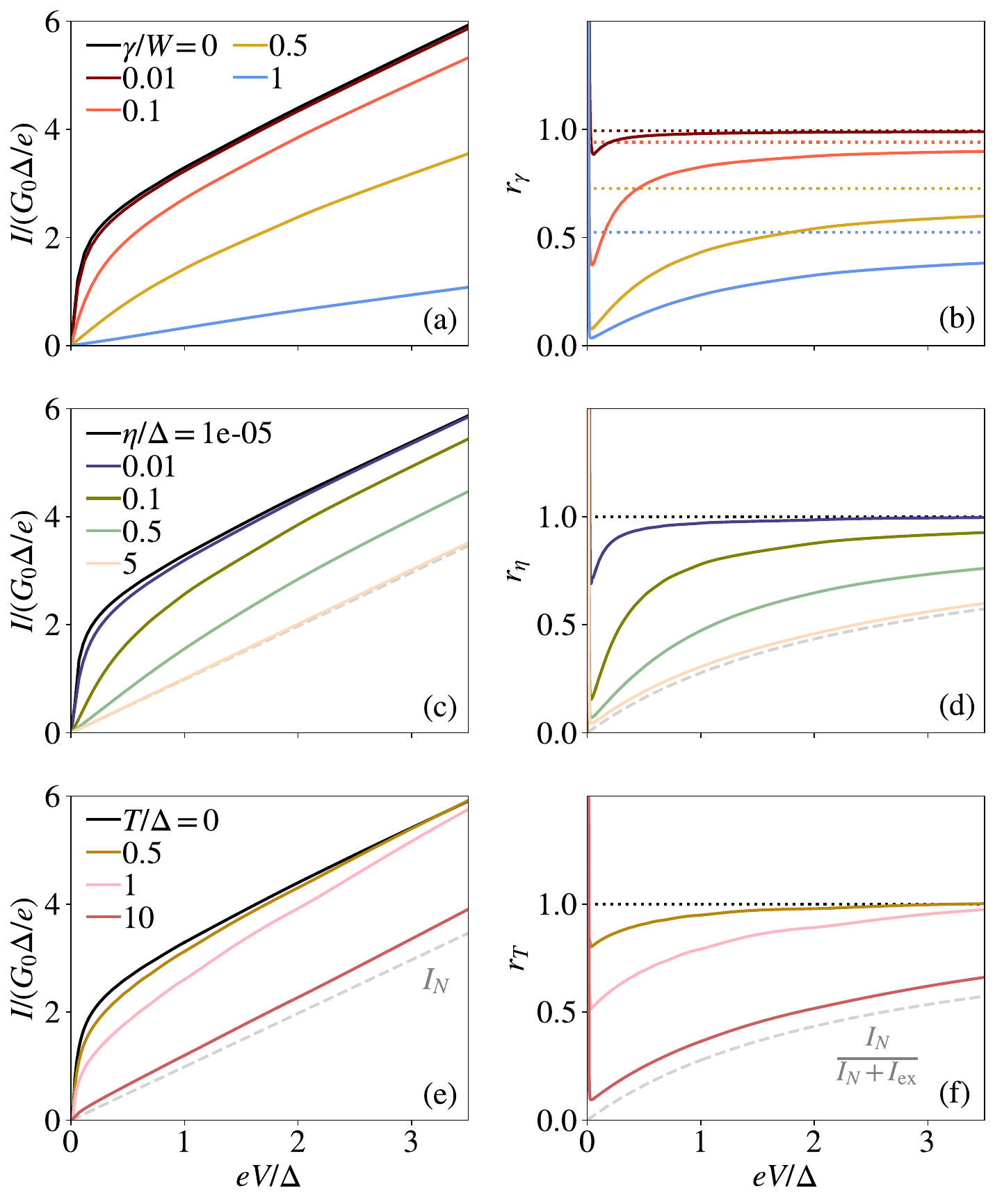}
\caption{(a) The conserved current $I$ as a function of voltage for different values of the dissipation strength $\gamma$, with $\tau/W = 0.9$. (b) The ratio of the currents with and without dissipation, $r_{\gamma} = I(\gamma > 0)/I(\gamma = 0)$ as a function of voltage for different values of $\gamma$. The solid lines indicate the superconducting reservoirs and the dotted lines the normal-state reservoirs. In panels (a) and (b), $\eta/\Delta = 10^{-5}$ and $T = 0$. (c) The current-voltage curves in the presence of a finite inelastic scattering rate $\eta$ with $\gamma = T = 0$ and (d) the corresponding ratio of the current at finite $\eta$ to the one with infinitesimal $\eta$, $r_{\eta} = I(\eta/\Delta > 10^{-5})/I(\eta/\Delta = 10^{-5})$. At large voltage, the current-voltage curves have the same slope and $r_{\eta} \to 1$. (e) The current-voltage curve at a finite temperature $T$ with $\gamma = 0$ and $\eta/\Delta = 10^{-5}$ and (f) the corresponding ratio $r_T = I(T > 0)/I(T = 0)$. The dashed gray lines in panels (c) and (e) indicate the linear normal-state current $I_N$. In panels (d) and (f), the dashed gray lines show the function $I_N/(I_N + I_{\text{ex}})$, where $I_{\text{ex}}$ is the excess current given by Eqs.~(\ref{eq:excess_current})--(\ref{eq:excess2}).}
\label{fig:high_transparency}
\end{figure}

We characterize transport through the dissipative quantum point contact by computing the conserved current $I$, given by Eq.~(\ref{eq:conserved_current}), as a function of voltage. Figure~\ref{fig:high_transparency} shows the current-voltage curves for different values of $\gamma$ and a high ($\tau/W = 0.9$) transparency of the contact. For voltages below $2\Delta/e$, the current is generated by multiple Andreev reflections, a high-order cotunneling process where multiple Cooper pairs tunnel simultaneously with a quasiparticle. As dissipative couplings destroy quantum coherence, one could expect such a process to be strongly suppressed by dissipation. Figure~\ref{fig:high_transparency} shows however that there is no sharp suppression of the current at $eV = 2\Delta$ at any dissipation strength, but the current is seemingly uniformly reduced over most of the voltage range. This behavior was observed experimentally in the high-transparency and low-voltage ($eV/\Delta \lesssim 0.05$) regime~\cite{HuangEsslinger2023}. 

We compare in Fig.~\ref{fig:high_transparency} the effect of a particle loss at the contacts to that of a finite inelastic scattering rate $\eta$ and a finite temperature in the reservoirs. The corresponding current-voltage characteristics are shown in panels~(c) and~(e), respectively. While the particle loss modifies the slope of the current-voltage curve, a finite $\eta$ or $T$ leads to a reduction of the current but with a slope that at $eV \gg 2\Delta$ is unchanged. In the $\eta/\Delta \gg 1$, $T/\Delta \gg 1$ limit, the curves approach the $I_N = G_0 \alpha V$ line for normal-state reservoirs.

In order to distinguish whether the dissipation, inelastic scattering, and finite temperature have a different effect at voltages below and above $2\Delta/e$, we compute the ratio of the current at nonzero and zero (infinitesimal for $\eta$) values of these parameters. The ratio of current in the presence of dissipation and without dissipation, $r_{\gamma} = I(\gamma > 0)/I(\gamma = 0)$, is computed setting $\eta = 10^{-5}$ and $T = 0$. It is shown in Fig.~\ref{fig:high_transparency}(b), where the solid lines indicate superconducting reservoirs. For reference, we also plot the corresponding ratio for normal-state reservoirs, indicated by the dotted lines. The normal-state current-voltage relation is linear with a slope modified by the dissipation, given by Eq.~(\ref{eq:transparency}), and therefore $r_{\gamma}$ does not depend on the voltage. For the superconducting reservoirs, on the contrary, the ratio of the currents has a voltage dependence. It approaches the normal-state value at large voltage, whereas at $eV \lesssim 2 \Delta$, the current is reduced more in proportion to the nondissipative case. This indicates that the dissipation has a destructive effect on the high-order multiple Andreev reflections generating the current at low voltages. 

The stronger suppression of current at small voltages occurs also in the case of a finite inelastic scattering rate, as shown in Fig.~\ref{fig:high_transparency}(d). The damping of high-order MAR processes due to inelastic scattering was discussed previously in Refs.~\cite{LevyYeyatiCuevas1996,cuevas_superconducting1996}, where for $\eta > eV$, the current was found to arise from low-order tunneling processes while high-order MAR are damped. For large transparencies, the undamped current at large voltages approaches the normal-state current shifted by the excess current, $I \to I_N + I_{\text{ex}}$, as discussed in Section~\ref{sec:excess_current}. The ratio $r_{\eta} = I(\eta/\Delta > 10^{-5})/I(\eta/\Delta = 10^{-5})$ therefore approaches the function $I_N/(I_N + I_{\text{ex}})$ for large inelastic scattering rates $\eta/\Delta \gg 1$ where the current approaches the normal-state value $I_N$. This limiting curve is shown by the dashed gray lines in panels~(d) and~(f). A finite temperature of the reservoirs similarly leads to a suppression of high-order MAR at small voltages, as seen in Fig.~\ref{fig:high_transparency}(f). At high temperature $T/\Delta \gg 1$, the current approaches the normal-state value $I_N$ which is independent of temperature.

In Figs.~\ref{fig:high_transparency}(b, d, f), we also observe an upturn of the curve at very small voltages $eV \lesssim (1 - \alpha) \Delta$. In the absence of losses or inelastic scattering, at zero temperature, the current is proportional to $(\tau/W)^{2n}$ where $n$ is the order of the multiple Andreev reflections generating the current~\cite{cuevas_superconducting1996}, or the number of Cooper pairs transmitted in the process $n \sim \Delta/V$. For $\tau/W < 1$, the current due to MAR is therefore suppressed at small voltages, and perturbations which allow lower-order MAR to contribute enhance the current in this regime. For high transparencies, the relevant voltage range is very small ($eV/\Delta \lesssim 0.01$) and the enhancement is not visible in Figs.~\ref{fig:high_transparency}(a, c, e). The enhancement of the current in the low-transparency regime is analyzed in more detail in Section~\ref{sec:weak_tunneling}.

\subsection{Excess current in the lossy quantum point contact}
\label{sec:excess_current}

\begin{figure}[h]
\includegraphics[width=0.49\linewidth]{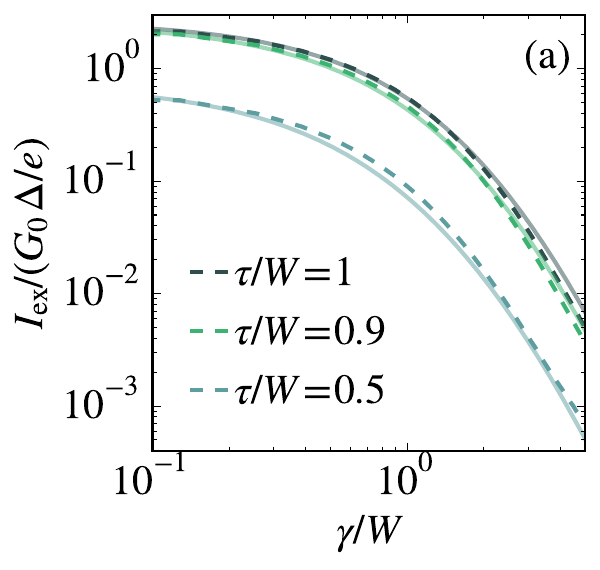}
\includegraphics[width=0.49\linewidth]{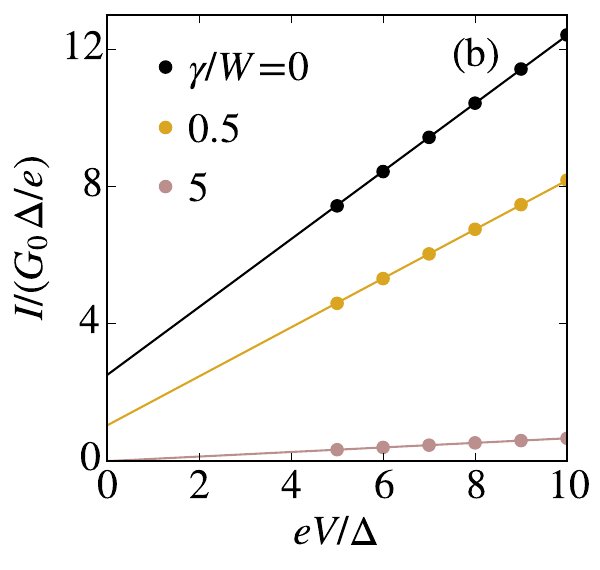}
\caption{(a) The excess current $I_{\text{ex}}$ as a function of the dissipation strength $\gamma$ for different tunneling amplitudes. The dashed lines show the numerical results and the solid lines show $I_{\text{ex}}$ given by Eqs.~(\ref{eq:excess_current})--(\ref{eq:excess2}). (b) The excess current is given numerically by the intercept of the lines $\alpha V + I_{\text{ex}}$ with the vertical axis. Here, the tunneling amplitude is $\tau/W = 1$.}
\label{fig:excess_current}
\end{figure}

For increasing voltage $eV \gg 2 \Delta$, the slope of the current-voltage curve approaches that of the noninteracting reservoirs, while the curves are shifted higher on the vertical axis. This shift is the excess current $I_{\text{ex}} = \lim_{eV/\Delta \to \infty} (I - I_{N})$ generated by the lowest-order pair cotunneling processes as the contribution of higher-order processes quickly decays at voltages larger than $2\Delta/e$.
In Fig.~\ref{fig:excess_current}(a), we plot the excess current as a function of the dissipation strength in the lossy quantum point contact. It is obtained by assuming a linear dependence $I = 2\alpha G_0 V + I_{\text{ex}}$ at large voltage $eV > 2\Delta$, with $\alpha$ given by the normal-state expression~(\ref{eq:transparency}). The constant $I_{\text{ex}}$ is solved from numerical data points by setting the squared residuals to zero, $\sum_i \left( I_i - \alpha V_i - I_{\text{ex}} \right)^2 = 0$. The slope $\alpha$ is set as the normal-state expression of Eq.~(\ref{eq:transparency}).
We use numerical data points $(V_i, I_i)$ with voltages in the interval $eV_i/\Delta \in [5, 10]$, as shown in Fig.~\ref{fig:excess_current}(b).

Figure~\ref{fig:excess_current} also shows the analytic expression 
\begin{equation}
I_{\text{ex}}(\alpha) = I_{\text{ex} 1}(\alpha) + I_{\text{ex} 2}(\alpha),
\label{eq:excess_current}
\end{equation} 
where
\begin{align}
I_{\text{ex} 1} &= \frac{\Delta}{h} \frac{\alpha^2}{(2 - \alpha) \sqrt{1 - \alpha}} 
\ln\left( \frac{1 + \frac{2\sqrt{1 - \alpha}}{2 - \alpha}}{1 - \frac{2\sqrt{1 - \alpha}}{2 - \alpha}} \right),
\label{eq:excess1}
\\
I_{\text{ex} 2} &= \frac{\Delta}{h} \left[ \frac{\alpha^2}{1 - \alpha} + \frac{\alpha^2 (2 - \alpha)}{2(1 - \alpha)^{\frac{3}{2}}} \ln\left( \frac{1 - \sqrt{1 - \alpha}}{1 + \sqrt{1 - \alpha}} \right)\right].
\label{eq:excess2}
\end{align}
These expressions are derived for the tunneling Hamiltonian of Eq.~(\ref{eq:hamiltonian}) in the absence of dissipation in Ref.~\cite{cuevas_superconducting1996}. The two terms $I_{\text{ex} 1}$ and $I_{\text{ex} 2}$ are the contributions of energies inside and outside the superconducting gap, respectively.
Figure~\ref{fig:excess_current} shows that Eqs.~(\ref{eq:excess_current})--(\ref{eq:excess2}) agree well with the numerical results when the generalized expression~(\ref{eq:transparency}) for the transparency $\alpha$ of the lossy quantum point contact is used. In the limit $\alpha \to 1$, the excess current has the value $I_{\text{ex}} = (8/3)G_0\Delta/e$~\cite{zaitsev1980, KlapwijkTinkham1982, cuevas_superconducting1996}.

\section{Enhancement of current in the weak-tunneling limit}
\label{sec:weak_tunneling}

When the tunneling amplitude is small, $\tau/W \ll 1$, the current is quickly reduced at voltages $eV < 2 \Delta$. The effect of the particle loss, inelastic scattering, and finite temperature is to reduce the current at large voltages, but they also lead to an enhancement of the current when the voltage is small. Here, we examine the weak-tunneling limit where the current in the absence of particle loss or inelastic scattering can be approximated by taking into account only the first-order quasiparticle tunneling~\cite{blonder_transition1982},
\begin{align}
\begin{split}
I \propto \int_{-\infty}^{\infty} d\omega \: &\rho \left( \omega - \mu_L \right) \rho\left( \omega - \mu_R \right) \\
&\times \left[ n_F\left(\omega-\mu_L\right) - n_F\left(\omega-\mu_R\right) \right].
\end{split}
\label{eq:fermi_golden_rule}
\end{align}
This formula is derived for the Hamiltonian system in the absence of dissipation and neglects the contribution of Andreev reflections. Here, $\rho(\omega)$ is the local density of states (DOS).
A finite particle loss rate or inelastic scattering rate lead to a modification of the local DOS, and in the following, we compare the full numerical solution to the weak-tunneling approximation~(\ref{eq:fermi_golden_rule}) using the modified local density of states.

\subsection{Local density of states}

\begin{figure}[h]
\includegraphics[width=\linewidth]{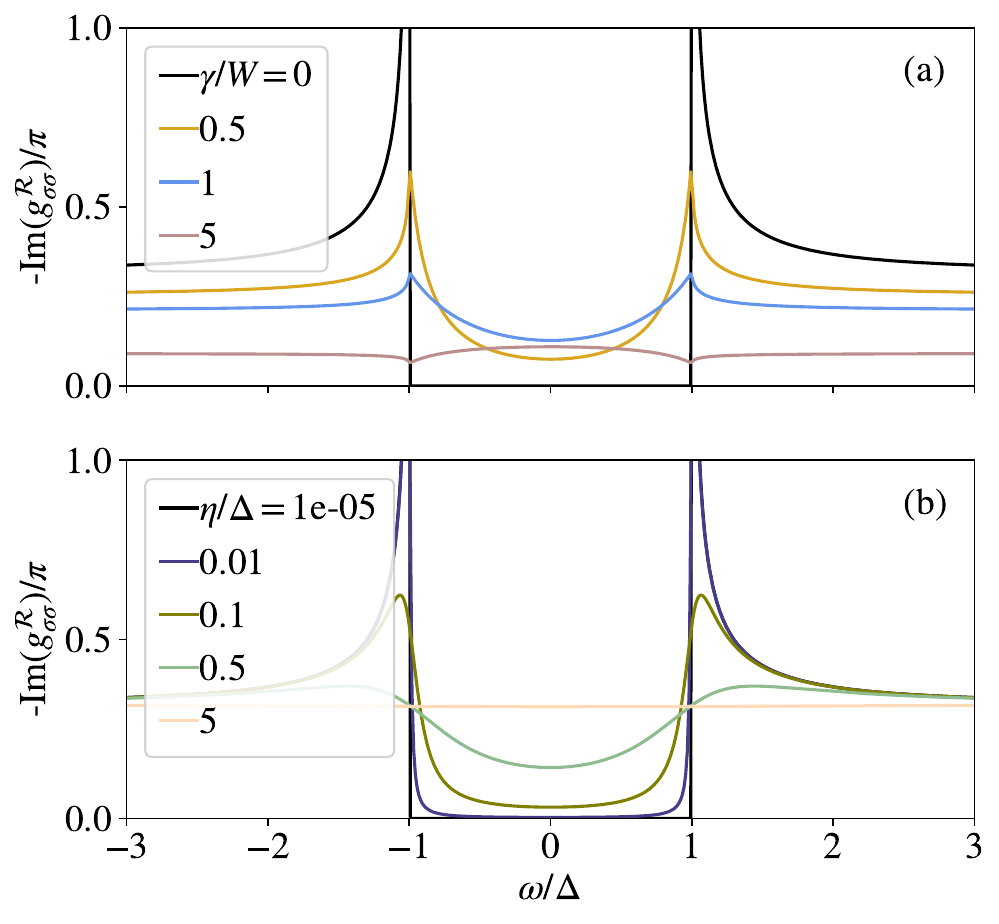}
\caption{(a) The local density of states as a function of frequency, given by Eqs.~(\ref{eq:local_dos}) and~(\ref{eq:normal_gf}), with $\eta/\Delta = 10^{-5}$. For large $\gamma$, the divergences are smoothened out and the local DOS becomes nearly uniform with a reduced value, approaching zero for $\gamma/W \to \infty$. (b) The local DOS for $\gamma = 0$ and finite quasiparticle damping rate $\eta$. For $\eta/\Delta \gg 1$, the divergences are smoothened out and the DOS approaches the normal-state value.}
\label{fig:local_dos}
\end{figure}

Both $\gamma$ and $\eta$ enter as an imaginary part in the retarded Green's function and a finite value leads to a broadening of the local density of states, connected to a finite lifetime of the quasiparticle excitations. Figure~\ref{fig:local_dos} shows the local DOS 
\begin{equation}
\rho(\omega) = -\frac{1}{\pi} \text{Im} [g_{\sigma\sigma}^\mathcal{R}(\mathbf{r} = \mathbf{0}, \omega)]
\label{eq:local_dos}
\end{equation}
of a BCS superconductor, corresponding to uncoupled reservoirs. When a local, spin-independent particle loss is applied at $\mathbf{r} = \mathbf{0}$, the normal Green's functions $g^\mathcal{R}_{\uparrow \uparrow}$ and $g^\mathcal{R}_{\downarrow \downarrow}$ become
\begin{align}
\begin{split}
g^\mathcal{R}_{\sigma \sigma}(\omega) = &\frac{1}{\text{det}\left([g^\mathcal{R}]^{-1}_{\sigma \sigma} \right)} \\
&\times \left( \frac{W(\omega + i \eta)}{\sqrt{\Delta^2 - (\omega + i \eta)^2}} + \frac{i \gamma}{2} \right),
\end{split}
\label{eq:normal_gf}
\end{align}
where 
\begin{equation*}
\text{det}\left([g^\mathcal{R}]^{-1}_{\sigma \sigma} \right) = -W^2 - \frac{\gamma^2}{4} + \frac{i W \gamma(\omega + i \eta)}{\sqrt{\Delta^2 - (\omega + i \eta)^2}}.
\end{equation*}
For $\gamma, \eta \to 0$, the local density of states reduces to $\rho(\omega) = \text{Re}\left(|\omega|/\sqrt{\omega^2 - \Delta^2} \right)$. 

In the $\gamma, \eta \to 0$ limit, the local DOS diverges at the gap edges. For increasing $\gamma$, the divergences are smoothened out and the magnitude of the local DOS outside the gap is reduced. The density of states gains finite values within the gap, which can be understood to result from pair breaking due to the particle loss leading to new quasiparticle states. Interestingly, the derivative of the local DOS remains discontinuous and for large dissipation $\gamma/W \gg 1$, the maxima at the gap edges are replaced by minima. A finite inelastic scattering rate similarly leads to a broadening of the divergences, while the maxima are smoothened out and shifted towards higher frequencies. In the $\eta/\Delta \to \infty$ limit, the local DOS approaches the constant value for normal-state reservoirs. A similar broadening of the BCS density of states was measured in a granular superconductor with enhanced inelastic scattering~\cite{DynesGarno1978, DynesOrlando1984}.

\subsection{Current-voltage characteristics}

We compare the current-voltage characteristics in the presence of particle loss, inelastic scattering, and a finite temperature to the single-quasiparticle tunneling approximation of Eq.~(\ref{eq:fermi_golden_rule}) in the limit of weak tunneling. This allows us to determine whether the current, in this limit, is generated by the tunneling of quasiparticles with a modified local DOS or whether higher-order tunneling processes contribute. 

\begin{figure}[t]
\includegraphics[width=\linewidth]{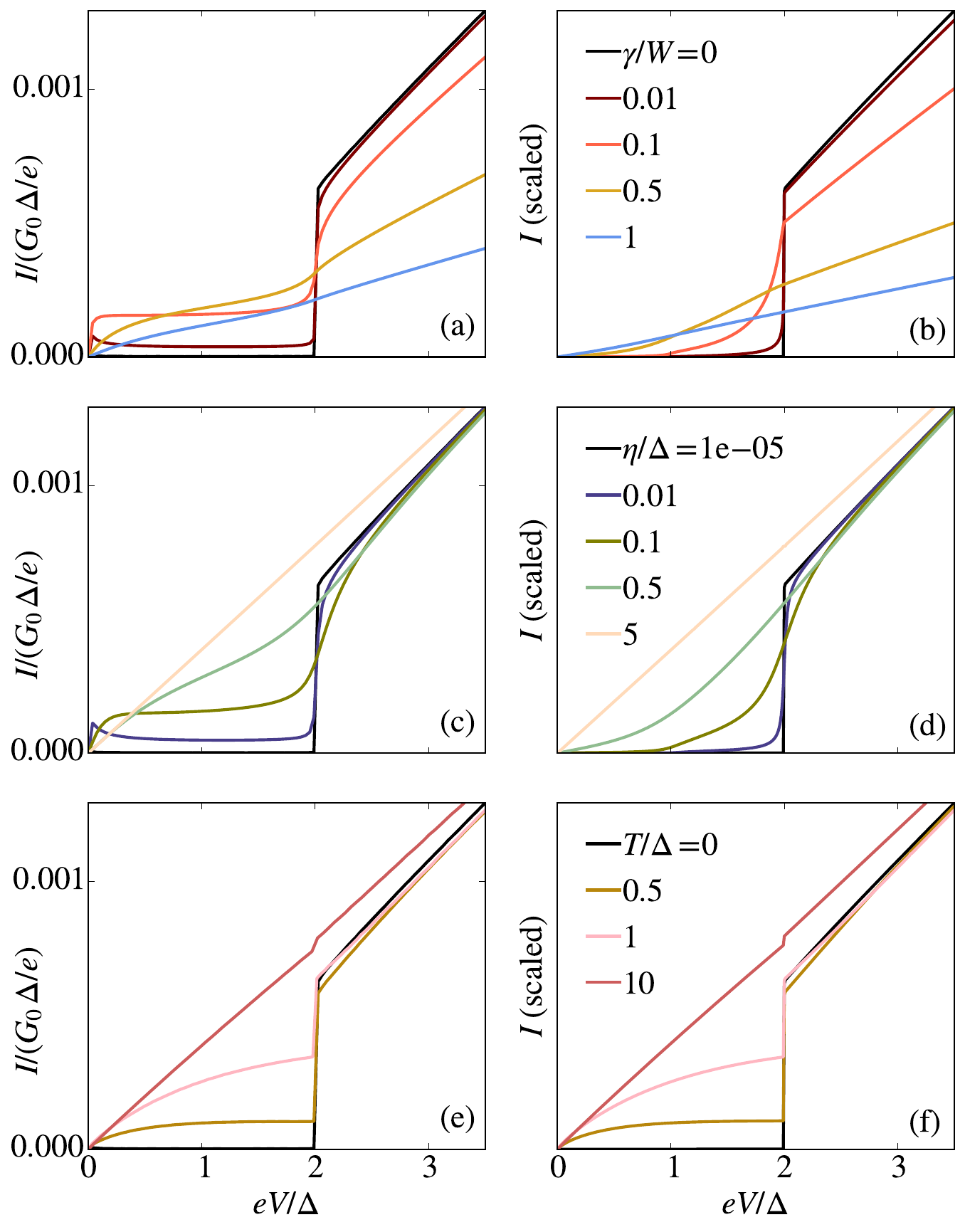}
\caption{(a) Current as a function of voltage for $\tau/W = 0.01$, $T = 0$, $\eta/\Delta = 10^{-5}$ and different dissipation strengths. (b) The current given by Eq.~(\ref{eq:fermi_golden_rule}), using the local density of states shown in Fig.~\ref{fig:local_dos}(a). (c) Current as a function of voltage in the presence of a finite quasiparticle damping rate $\eta$. Here, $\gamma = T = 0$. (d) The current given by Eq.~(\ref{eq:fermi_golden_rule}), using the local density of states of Fig.~\ref{fig:local_dos}(b). (e) Current as a function of voltage for varying temperature, with $\gamma = 0$ and $\eta/\Delta = 10^{-5}$. (f) The first order tunneling approximation of Eq.~(\ref{eq:fermi_golden_rule}) agrees well with the finite-temperature curves.}
\label{fig:weak_tunneling}
\end{figure}

In Fig.~\ref{fig:weak_tunneling}(a), the current as a function of voltage is shown for $\tau/W = 0.01$ and varying loss rate $\gamma$. In the absence of particle loss, the current is close to zero for $eV < 2\Delta$ and is reproduced by Eq.~(\ref{eq:fermi_golden_rule}), shown in Fig.~\ref{fig:weak_tunneling}(b). An increasing loss rate leads first to an enhancement of current at $eV < 2\Delta$ and then a reduction, while for $eV > 2\Delta$, the current is reduced by the loss. These features are present both in the full numerical solution and in the approximation~(\ref{eq:fermi_golden_rule}), which is computed using the modified local DOS of Eqs.~(\ref{eq:local_dos}) and~(\ref{eq:normal_gf}). The full solution, however, develops an additional step at $V \approx 0$, which appears at small dissipation strength $\gamma/W \lesssim 0.1$ and is smoothened out for stronger dissipation. This additional step is not reproduced by the first-order tunneling approximation~(\ref{eq:fermi_golden_rule}), which indicates that it originates from the MAR processes. We find that the calculation of the current converges already when only the lowest-order pair tunneling is accounted for. The presence of dissipation, due to the broadened density of states, seems to therefore allow a current to be generated through low-order MAR processes in the regime where the current is otherwise suppressed. 
For large $\gamma$, the DOS becomes independent of frequency. This uniform density of states resembles that of a system in the normal state, which is consistent with the current-voltage curves becoming linear for sufficiently large $\gamma$.

In the case of a finite inelastic scattering rate, the current is similarly enhanced for $eV < 2\Delta$, as seen in Fig.~\ref{fig:weak_tunneling}(c). It develops a step at small voltage and small~$\eta$, $eV \approx \eta/\Delta < 1$, which is smoothened out for $\eta/\Delta \gtrsim 1$. The first-order tunneling approximation~(\ref{eq:fermi_golden_rule}) is shown in Fig.~\ref{fig:weak_tunneling}(d). It again displays an enhancement of current in the small-voltage regime, due to the finite value of the local DOS within the gap, but does not reproduce the step at $eV \approx \eta/\Delta$. Unlike a large particle loss which leads to a vanishing current, a large inelastic scattering rate $\eta/\Delta \gg 1$ leads to a current-voltage curve that approaches the linear dependence $I_N = \alpha G_0 V$ of normal-state reservoirs.

The experimentally relevant situation of a finite temperature in the reservoirs produces a finite current at $eV < 2\Delta$, as quasiparticles are thermally excited to states above the gap. The resulting voltage dependence is qualitatively different from the lossy contact and the 
finite inelastic scattering rate in the reservoirs. The finite-temperature current-voltage curves are shown in Figure~\ref{fig:weak_tunneling}(e), where $\gamma = 0$, $\eta/\Delta = 10^{-5}$, and the temperature is varied while the superconducting gap $\Delta$ is constant. Compared to the local dissipation in Fig.~\ref{fig:weak_tunneling}(a), the finite temperature leads to a monotonic increase of current at low voltages and does not produce a nonmonotonic behavior as there is no overall loss of particles. Since there is no modification of the local DOS but only changes in the occupation of the states, the original step-like structure is retained at nonzero temperatures. This is seen more clearly for intermediate transparencies of the contact, discussed in Appendix~\ref{app:intermediate}. Most notably, the additional step at $V \approx 0$ which arises in the presence of dissipation is absent in the finite-temperature curves. The current-voltage curves are well reproduced by the first-order tunneling approximation~(\ref{eq:fermi_golden_rule}) shown in Fig.~\ref{fig:weak_tunneling}(f), indicating that the enhancement of the current originates from first-order quasiparticle tunneling.

\section{Discussion and conclusions}
\label{sec:conclusions}

We study transport through a lossy quantum point contact between two superconductors, modeled by a local tunneling between two superconducting reservoirs and a local particle loss at the contacts. We characterize transport by computing the conserved current as a function of voltage, using the Keldysh-Schwinger formalism. For highly transparent contacts, we find that the current is reduced by the dissipation. The slope of the current-voltage curve approaches the one of normal-state reservoirs coupled by a lossy quantum point contact at large voltages $eV \gg 2\Delta$. The excess current in the lossy system is reproduced by an analytic formula for a quantum point contact without losses~\cite{cuevas_superconducting1996}, when the modification of the transparency by the loss is taken into account.
While a large particle loss leads to a vanishing current, strong inelastic scattering or a high temperature in the reservoirs reproduce the normal-state current-voltage curve.

In the small-voltage regime $eV < 2\Delta$, the current depends exponentially on the order of the multiple Andreev reflections generating the current. It is therefore suppressed at small voltages $eV \lesssim (1 - \alpha)\Delta$ where the order is high. In this regime, the particle loss leads to an enhancement of the current as the local density of states is broadened and gains finite values within the gap, allowing lower-order MAR to contribute. The current-voltage curve develops a step for $0< \gamma/W < 1$ which is not reproduced by single-quasiparticle tunneling. We attribute this feature to a combined effect of the modified local DOS and the low-order multiple Andreev reflections, although we do not have a complete microscopic picture of this process. Here, calculating the shot noise could give information on the number of transmitted charge carriers~\cite{CuevasLevyYeyati1999}. Weak inelastic scattering produces a similar step-like feature. While the current at $eV < 2\Delta$ has a nonmonotonic dependence on the particle loss, for strong inelastic scattering, the current-voltage curve approaches the normal-state result. A finite temperature leads to an enhancement of the current due to thermal excitations and, unlike in the case of particle loss or inelastic scattering, is well described by single-quasiparticle tunneling. 

The theoretical predictions for the lossy contact agree with experimental data in the high-transparency and low-voltage regime~\cite{HuangEsslinger2023}. The predictions for the low-transparency regime, specifically the enhancement of current with dissipation, could be studied experimentally in a similar setup with reduced transparency of the contact~\cite{yao_controlled_2018}. 
There are several mechanisms at play which may contribute to the observed current-voltage curves: Besides the reduction of current by removal of particles, the accompanying decoherence can be destructive to the high-order coherent tunneling process. Comparing the effect of the single-particle loss to that of a pair loss, or to a pure dephasing, could help to characterize the physical mechanism responsible for the current in this system.

\begin{acknowledgments}
We gratefully acknowledge P. Fabritius, S. H\"ausler, M. Talebi, and S. Wili for discussions. A.-M.V. acknowledges funding from the Deutsche Forschungsgemeinschaft (DFG, German Research Foundation) in particular under project number 277625399 - TRR 185 (B3) and project number 277146847 - CRC 1238 (C05) and Germany’s Excellence Strategy -- Cluster of Excellence Matter and Light for Quantum Computing (ML4Q) EXC2004/1 -- 390534769. S.U. acknowledges MEXT Leading Initiative for Excellent Young Researchers
 (Grant No. JPMXS0320200002), JSPS KAKENHI (Grant No.~JP21K03436), and Matsuo Foundation. J.M., M.H., and T.E. acknowledge the Swiss National Science Foundation (Grants No.~182650, No.~212168 and No.~NCCR-QSIT) and European Research Council advanced grant TransQ (Grant No.~742579) for funding. This work was supported in part by the Swiss National Science Foundation under Division II (Grant 2000020-188687).
\end{acknowledgments}

\appendix

\section{Currents}
\label{app:currents}

For an open quantum system described by the master equation~(\ref{eq:master_equation}), the time derivative of the particle number is obtained as
\begin{align}
\frac{d}{dt} \braket{N_i} &= \frac{d}{dt} \text{Tr}\left( N_i \rho(t) \right) \nonumber \\
&= -i\text{Tr}\left( N_i [H, \rho] \right) \nonumber \\
&\: + \gamma \text{Tr} \left( 
N_i \psi_i(\mathbf{0}) \rho \psi_i^{\dagger}(\mathbf{0}) - \frac{1}{2} N_i \left\{ \psi_i^{\dagger}(\mathbf{0}) \psi_i(\mathbf{0}), \rho \right\}
\right) \nonumber \\
&= -i \braket{[N_i, H]} -\gamma \braket{\psi_i^{\dagger}(\mathbf{0}) \psi_i(\mathbf{0})}
\label{eq:time_derivative}
\end{align}
where $H$ is the Hamiltonian of Eq.~(\ref{eq:hamiltonian}) and $i = L, R$. We have left out the spin index here as the equations are the same for both spins. The conserved current $I$ of Eq.~(\ref{eq:conserved_current}) is given by the first term,
\begin{align}
I &= i \sum_{\sigma = \uparrow, \downarrow} \braket{[N_{L \sigma}, H]} \\
&= i \sum_{\sigma = \uparrow, \downarrow} \left( \braket{[N_{L \sigma}, H_{\text{tun}}]} + i \braket{[N_{L \sigma}, H_L} \right),
\label{eq:tunneling_and_bcs}
\end{align}
where $\braket{[N_{L \sigma}, H_{\text{tun}}]} = -\braket{[N_{R \sigma}, H_{\text{tun}}]}$ and $H_L$ is the BCS Hamiltonian of Eq.~(\ref{eq:reservoir_hamiltonian}). Writing $H_i$ in position basis, we find the second term in Eq.~(\ref{eq:tunneling_and_bcs}) as
\begin{align*}
\begin{split}
&\braket{[N_{i \sigma}, H_i]} \\
&= \Delta \left\langle \left[ \int d\mathbf{r} \psi_{i \sigma}^{\dagger}(\mathbf{r}) \psi_{i \sigma}^{\phantom{\dagger}}(\mathbf{r}), \int d\mathbf{r'} \psi_{i \uparrow}^{\dagger}(\mathbf{r'}) \psi_{i \downarrow}^{\dagger}(\mathbf{r'}) \right] \right \rangle \\
&\: + \Delta \left\langle \left[\int d\mathbf{r} \psi_{i \sigma}^{\dagger}(\mathbf{r}) \psi_{i \sigma}^{\phantom{\dagger}}(\mathbf{r}), \int d\mathbf{r'} \psi_{i \downarrow}(\mathbf{r'}) \psi_{i \uparrow}(\mathbf{r'}) \right] \right \rangle \\
&= \Delta \int d\mathbf{r} \left(\left\langle \psi_{i \uparrow}^{\dagger}(\mathbf{r}) \psi_{i \downarrow}^{\dagger}(\mathbf{r})\right\rangle - \left \langle \psi_{i \downarrow}(\mathbf{r}) \psi_{i \uparrow}(\mathbf{r}) \right \rangle \right) = 0.
\end{split}
\end{align*}
Here, we have identified the real-valued order parameter $\Delta = V \left\langle \psi_{L \uparrow}^{\dagger}(\mathbf{r}) \psi_{L \downarrow}^{\dagger}(\mathbf{r})\right\rangle = V \left \langle \psi_{L \downarrow}(\mathbf{r}) \psi_{L \uparrow}(\mathbf{r}) \right \rangle$, where $V$ is the contact interaction, so that the terms on the last line cancel.
The second term in Eq.~(\ref{eq:time_derivative}) is equal to the loss current $I_i^{\text{loss}} = -\gamma \sum_{\sigma = \uparrow, \downarrow} \braket{\psi_{i \sigma}^{\dagger}(\mathbf{0}) \psi_{i \sigma}^{\phantom{\dagger}}(\mathbf{0})}$.

\section{Keldysh action for the dissipative system}
\label{app:Keldysh_action}

The Keldysh action for the dissipative system is written as~\cite{Sieberer_Keldysh2016}
\begin{equation*}
S = \int_{-\infty}^{\infty} dt \left[ \bar{\psi}^+ \partial_t \psi^+ - \bar{\psi}^- \partial_t \psi^- - i \mathcal{L}(\bar{\psi}^+, \psi^+, \bar{\psi}^-, \psi^-) \right].
\end{equation*}
For uncoupled reservoirs, $\psi^{\pm}$ denotes the vector $\psi_i^{\pm} = (\psi_{i \uparrow}^{\pm} \; \bar{\psi}_{i \downarrow}^{\pm})^T$ with $i = L, R$. The Lindblad term is of the form
\begin{align}
\begin{split}
\mathcal{L} (&\bar{\psi}^+, \psi^+, \bar{\psi}^-, \psi^-) = -i (H^+ - H^-) \\
+ &\gamma \sum_{a} \left( \bar{L}_a^- L_a^+ - \frac{1}{2} \bar{L}_a^+ L_a^+ 
- \frac{1}{2} \bar{L}_{\alpha}^- L_{\alpha}^- \right),
\end{split}
\label{eq:lindblad_term}
\end{align}
where $H^{\pm}$ is a function of $\psi^{\pm}$, and $a$ denotes arbitrary degrees of freedom. In the case of a local particle loss from the reservoirs, $L_a = \psi_{i \sigma}(\mathbf{0})$, and we can identify
\begin{align}
\begin{split}
S_{\text{loss}} = -i \gamma &\int_{-\infty}^{\infty} \frac{d \omega}{2 \pi} \sum_{i = L, R} \sum_{\sigma = \uparrow, \downarrow} \Big[ \bar{\psi}_{i \sigma}^-(\mathbf{0}) \psi_{i \sigma}^+(\mathbf{0}) \\
&- \frac{1}{2} \bar{\psi}_{i \sigma}^+(\mathbf{0}) \psi_{i \sigma}^+(\mathbf{0}) - \frac{1}{2} \bar{\psi}_{i \sigma}^-(\mathbf{0}) \psi_{i \sigma}^-(\mathbf{0}) \Big].
\end{split}
\end{align}

We use here the frequency representation. Since the interaction term couples spin-up and spin-down fermions at opposite relative frequencies with respect to the chemical potential, it is convenient to use the relative frequency argument $\bar{\omega} = \omega - \mu$. We then have
\begin{equation}
\psi_i^{\pm}(\mathbf{0}, \bar{\omega}) = 
\begin{pmatrix}
\psi_{i \uparrow}^{\pm}(\mathbf{0}, \bar{\omega}) \\
\bar{\psi}_{i \downarrow}^{\pm}(\mathbf{0}, -\bar{\omega})
\end{pmatrix}.
\end{equation}
Rotating the variables as
\begin{align*}
&\psi^+ = \frac{1}{\sqrt{2}} \left( \psi^1 + \psi^2 \right), \hspace{5mm}
&\psi^- = \frac{1}{\sqrt{2}} \left( \psi^1 - \psi^2 \right),
\end{align*}
we obtain the standard matrix for for $S_{\text{loss}} = S_{\text{loss}, \uparrow} + S_{\text{loss}, \downarrow}$. In the frequency representation,
\begin{align*}
S_{\text{loss}, \uparrow} = -i \gamma \sum_{i = L, R} \int_{-\infty}^{\infty} \frac{d \omega}{2 \pi} &
\begin{pmatrix}
\bar{\psi}_{i \uparrow}^1(\bar{\omega}) &\bar{\psi}_{i \uparrow}^2(\bar{\omega})
\end{pmatrix} \\
&\times
\begin{pmatrix}
0	&-\frac{i \gamma}{2} \\
\frac{i \gamma}{2}	&i\gamma
\end{pmatrix}
\begin{pmatrix}
\psi_{i \uparrow}^1(\bar{\omega})	\\
\psi_{i \uparrow}^2(\bar{\omega})
\end{pmatrix}
\end{align*}
and
\begin{align*}
S_{\text{loss}, \downarrow} = -i \gamma \sum_{i = L, R} \int_{-\infty}^{\infty} \frac{d \omega}{2 \pi} &
\begin{pmatrix}
\psi_{i \downarrow}^1(-\bar{\omega}) &\psi_{i \downarrow}^2(-\bar{\omega})
\end{pmatrix} \\
&\times
\begin{pmatrix}
0	&-\frac{i \gamma}{2} \\
\frac{i \gamma}{2}	&-i\gamma
\end{pmatrix}
\begin{pmatrix}
\bar{\psi}_{i \downarrow}^1(-\bar{\omega})	\\
\bar{\psi}_{i \downarrow}^2(-\bar{\omega})
\end{pmatrix}.
\end{align*}

\section{Action for the superconducting contact}
\label{app:infinite_matrix}

As discussed in Section~\ref{sec:superconductor-superconductor}, the presence of the mean-field interaction and a chemical potential bias leads to an action which is not block diagonal in frequency. Instead, it has matrix elements which couple fermions at increasing discrete frequencies given by Eq.~(\ref{eq:recursion}). To construct $\mathcal{G}^{-1}$ of Eq.~(\ref{eq:general_action}) for the two superconducting leads coupled by a tunneling term, we use a reorganized basis where we group the $\psi$s into blocks of spin up and spin down: 
\begin{widetext}
\begin{equation*}
\mathbf{\Psi}(\omega_0) = 
\begin{pmatrix}
\psi_{L \uparrow}^1(\bar{\omega}_0) &\psi_{L \uparrow}^2(\bar{\omega}_0) &\psi_{R \uparrow}^1(\bar{\omega}_0) &\psi_{R \uparrow}^2(\bar{\omega}_0) 
&\bar{\psi}_{L \downarrow}^1(\bar{\omega}_1) &\bar{\psi}_{L \downarrow}^2(\bar{\omega}_1)
&\bar{\psi}_{R \downarrow}^1(\bar{\omega}_1) &\bar{\psi}_{R \downarrow}^2(\bar{\omega}_1)
&\bar{\psi}_{R \downarrow}^1(\bar{\omega}_{-1}) &\bar{\psi}_{R \downarrow}^2(\bar{\omega}_{-1}) &\dots
\end{pmatrix}^T.
\end{equation*}
The order of the spinor elements can be chosen arbitrarily; it determines the structure of the matrix $\mathcal{G}^{-1}$ to be inverted but does not affect the resulting correlation functions. We denote the frequency relative to the chemical potential by $\bar{\omega}_n = \omega_n - \mu$, where the frequencies $\omega_n$ are given by Eq.~(\ref{eq:recursion}). The relative frequency in either lead is denoted by $\bar{\omega}_n$: For $\psi_{L \sigma}^{1, 2}(\bar{\omega}_n)$, $\bar{\omega}_n = \omega_n - \mu_L$ and for $\psi_{R \sigma}^{1, 2}(\bar{\omega}_n)$, $\bar{\omega}_n = \omega_n - \mu_R$. In this basis, $\mathcal{G}^{-1}$ has the structure
\begin{equation}
\mathcal{G}^{-1} =
\begin{pmatrix}
\mathbf{\Omega}_{L \uparrow}(\bar{\omega}_0)	&\mathcal{T}	&\mathbf{\Delta}_L(\bar{\omega}_0)	&0		&0	&\dots	&	&	&0 \\
\mathcal{T}			&\mathbf{\Omega}_{R \uparrow}(\bar{\omega}_0)&0				&0		&\mathbf{\Delta}_R(\bar{\omega}_0)	\\
\mathbf{\Delta}_L(\bar{\omega}_0)	&0		&\mathbf{\Omega}_{L \downarrow}(\bar{\omega}_1)	&-\mathcal{T}	&0		\\
0					&0		&-\mathcal{T}	&\mathbf{\Omega}_{R \downarrow}(\bar{\omega}_1)	&0	&0	&\mathbf{\Delta}_R(\bar{\omega}_2) \\
0					&\mathbf{\Delta}_R(\bar{\omega}_0)	&0			&0		&\mathbf{\Omega}_{R \downarrow}(\bar{\omega}_{-1}) &-\mathcal{T}	&0	& \\
\vdots	&		&			&0			&-\mathcal{T}	&\mathbf{\Omega}_{L \downarrow}(\bar{\omega}_{-1}) 	&0	&0	&\mathbf{\Delta}_L(\bar{\omega}_{-2}) \\
	&		&					&\mathbf{\Delta}_R(\bar{\omega}_2)		&0	&0		&\mathbf{\Omega}_{R \uparrow}(\bar{\omega}_2)	&\mathcal{T}	&0 \\
	&	&	&	&	&0	&\mathcal{T}	&\mathbf{\Omega}_{L \uparrow}(\bar{\omega}_2)	&\mathcal{T} \\
0	&	&	&	&	&\mathbf{\Delta}_L(\bar{\omega}_{-2})	&0	&\mathcal{T}	&\ddots
\end{pmatrix},
\label{eq:infinite_matrix_qpc}
\end{equation}
\end{widetext}
where the elements are matrices of size $2 \times 2$. The $\mathbf{\Omega}$ and $\mathbf{\Delta}$ blocks consist of the elements $\left[ g^{\mathcal{R}, \mathcal{A}, \mathcal{K}}_i(\bar{\omega}) \right]^{-1}_{m, n}$ of Eq.~(\ref{eq:retarded_advanced})--(\ref{eq:keldysh_block_dissipation}),
\begin{align}
\begin{split}
\mathbf{\Omega}_{i \uparrow}(\bar{\omega}) &= 
\begin{pmatrix}
0	&\left[ g^\mathcal{A}_i(\bar{\omega}) \right]^{-1}_{1, 1} \\
\left[ g^\mathcal{R}_i(\bar{\omega}) \right]^{-1}_{1, 1}	&\left[ \mathcal{G}^{-1}_i(\bar{\omega}) \right]^{\mathcal{K}}_{1, 1}
\end{pmatrix} \\
&= 
\begin{pmatrix}
0	&\left[ g^\mathcal{A}_i(\bar{\omega}) \right]^{-1}_{2, 2} \\
\left[ g^\mathcal{R}_i(\bar{\omega}) \right]^{-1}_{2, 2}	&\left[ \mathcal{G}^{-1}_i(\bar{\omega}) \right]^{\mathcal{K}}_{2, 2}
\end{pmatrix} = \mathbf{\Omega}_{i \downarrow}(-\bar{\omega}),
\end{split}
\label{eq:omega_block} \\
\begin{split}
\mathbf{\Delta}_i(\bar{\omega}) &= 
\begin{pmatrix}
0	&\left[ g^\mathcal{A}_i(\bar{\omega}) \right]^{-1}_{1, 2} \\
\left[ g^\mathcal{R}_i(\bar{\omega}) \right]^{-1}_{1, 2}	&\left[ \mathcal{G}^{-1}_i(\bar{\omega}) \right]^{\mathcal{K}}_{1, 2}
\end{pmatrix} \\
&= 
\begin{pmatrix}
0	&\left[ g^\mathcal{A}_i(\bar{\omega}) \right]^{-1}_{2, 1} \\
\left[ g^\mathcal{R}_i(\bar{\omega}) \right]^{-1}_{2, 1}	&\left[ \mathcal{G}^{-1}_i(\bar{\omega}) \right]^{\mathcal{K}}_{2, 1}
\end{pmatrix},
\end{split}
\end{align}
and the tunneling matrix block has the form
\begin{equation*}
\mathcal{T} = 
\begin{pmatrix}
0	&-\tau \\
-\tau	&0
\end{pmatrix}.
\end{equation*}

\begin{figure}[h]
\includegraphics[width=0.8\linewidth]{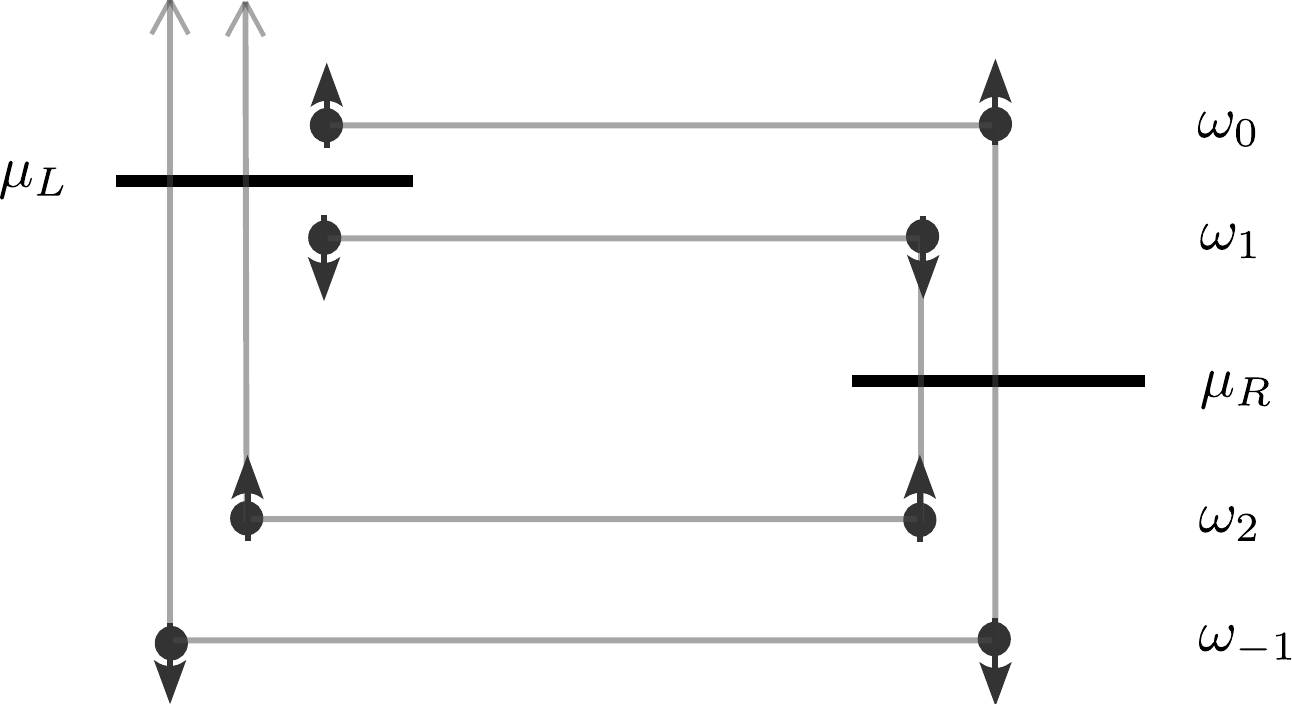}
\caption{Fermions with opposite spins are coupled by the interaction at opposite frequencies with respect to the chemical potential on either side. On the other hand, the tunneling term couples fermions across the contact at the same absolute frequency. This leads to the coupling of matrix elements at increasing frequency and the non-block-diagonal structure of the matrices in Eq.~(\ref{eq:infinite_matrix_qpc}), which reflects the multiple Andreev reflection or pair cotunneling process.}
\label{fig:hierarchy}
\end{figure}

The argument values $\bar{\omega}$ are connected so that the fermions with opposite spin which are coupled by the interaction on each side have the opposite relative frequencies: $\bar{\omega}_1 = -\bar{\omega}_0$ on the left and $\bar{\omega}_{-1} = -\bar{\omega}_0$ on the right, $\bar{\omega}_2 = -\bar{\omega}_1$ on the right and $\bar{\omega}_{-2} = -\bar{\omega}_{-1}$ on the left, etc. This ``hierarchy'' of frequencies~\cite{Bolech_point_contact2004,Bolech_Keldysh_study2005} is illustrated in Fig.~\ref{fig:hierarchy}. The matrix block for spin down in Eq.~(\ref{eq:omega_block}) is $\mathbf{\Omega}_{i \downarrow}(-\bar{\omega}) = \mathbf{\Omega}_{i \uparrow}(\bar{\omega})$
since the negative sign in the argument cancels with the one which comes from anticommuting $\psi$ and $\bar{\psi}$ in the Nambu-Keldysh spinor.

\section{Intermediate transparencies}
\label{app:intermediate}

\begin{figure}[h]
\includegraphics[width=\linewidth]{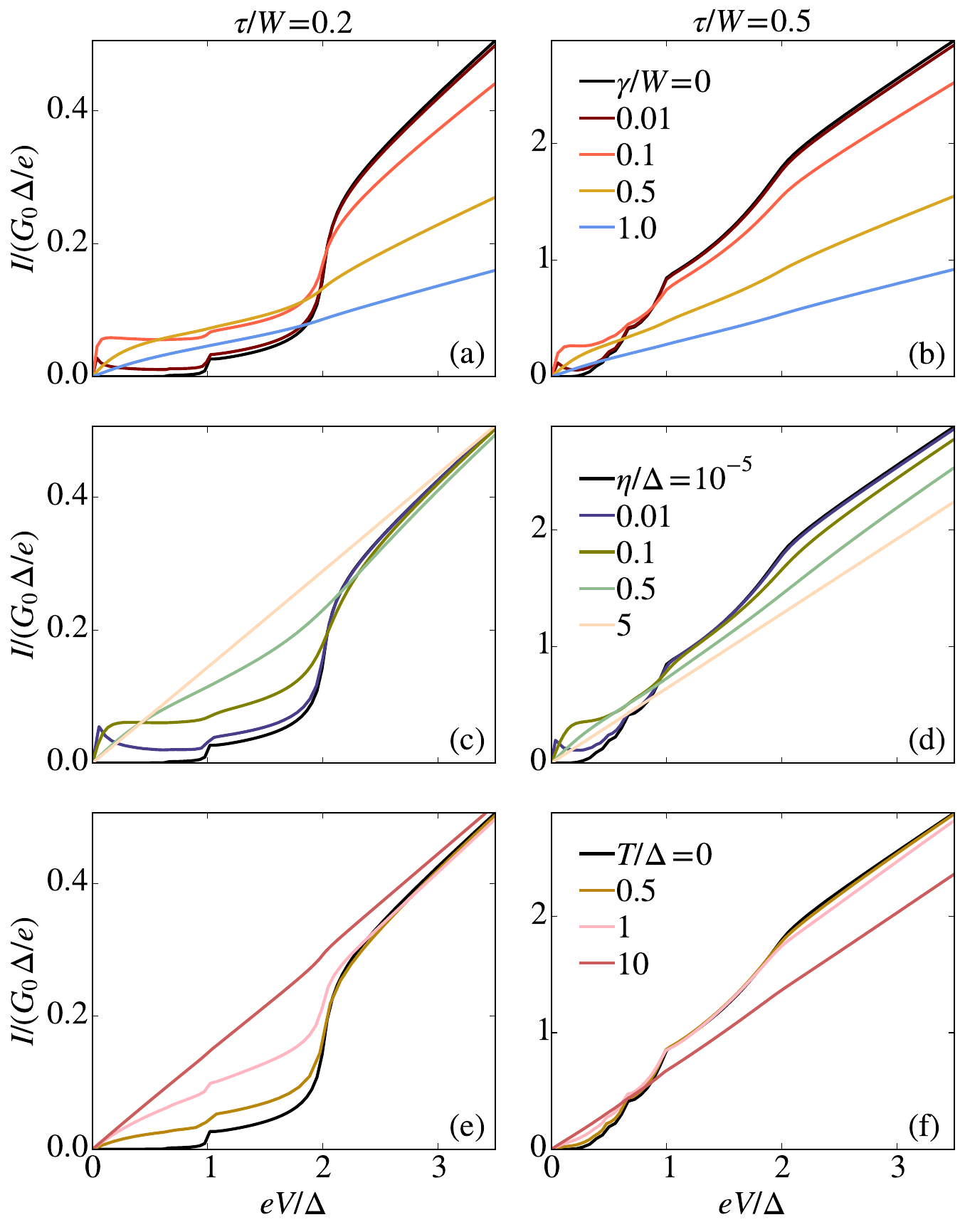}
\caption{Current as a function of voltage for (a, c, e) $\tau/W = 0.2$ and (b, d, f) $\tau/W = 0.5$. In panels (a, b), the loss rate $\gamma$ varies while $T = 0$, $\eta/\Delta = 10^{-5}$. Panels (c, d) correspond to a varying inelastic scattering rate $\eta$ with $\gamma = T = 0$. In (e, f), the temperature $T$ varies while $\gamma = 0$, $\eta/\Delta = 10^{-5}$.}
\label{fig:intermediate}
\end{figure}

Figure~\ref{fig:intermediate} shows the current-voltage curves for intermediate values of the tunneling amplitude, $\tau/W = 0.2$ and $\tau/W = 0.5$, similar to Figs.~\ref{fig:high_transparency} and~\ref{fig:weak_tunneling}. The current is enhanced by particle loss, inelastic scattering, and a finite temperature at small voltages $eV \lesssim 2\Delta$. At larger voltages, the effect of the particle loss is to reduce the current. On the other hand, the inelastic scattering and finite temperature in the reservoirs lead to a current-voltage curve that approaches the linear dependence of normal-state reservoirs. The current at $eV \gtrsim 2\Delta$ is therefore slightly enhanced for $\tau/W = 0.2$ and reduced for $\tau = 0.5$. Compared to the finite temperature, the steps in the current-voltage curve are more strongly smoothened out in the case of $\gamma > 0$ and $\eta > 0$ due to the broadening of the local density of states shown in Fig.~\ref{fig:local_dos}.

\bibliography{paper}

\end{document}